\documentclass{article}[12]
\usepackage{epsfig,psfig,float,amssymb,latexsym}
\usepackage[notref,notcite]{}
\newcommand{\NP}[1]{ Nucl.\ Phys.\ {\bf #1}}

\newcommand{\PL}[1]{ Phys.\ Lett.\ {\bf #1}}

\newcommand{\PR}[1]{Phys.\ Rev.\ {\bf #1}}

\newcommand{\bi}{\bibitem}
\newcommand{\vs}{\vspace{-0.20cm}}
\newcommand{\be}{\begin{equation}}
\newcommand{\ee}{\end{equation}}
\newcommand{\ba}{\begin{eqnarray}}
\newcommand{\ea}{\end{eqnarray}}

\newcommand{\nn}{\nonumber}

\begin{document}

\begin{center}
\huge{The $\rho$ Meson in a Nuclear Medium}
\end{center}
\begin{center}
{\Large{D. Cabrera, E. Oset, M. J. Vicente-Vacas}
\vspace{0.3cm}

\small{Departamento de F\'{\i}sica Te\'orica e IFIC, Centro Mixto
Universidad de Valencia-CSIC,\\
46100 Burjassot (Valencia), Spain}}
\end{center}
\begin{abstract}
{\small{In this work, propagation properties of the $\rho$ meson in
symmetric nuclear matter are studied. We make use of a coupled channel unitary approach to
meson-meson scattering, calculated from the lowest
order Chiral Perturbation Theory ($\chi PT$) lagrangian including explicit resonance
fields. Low energy chiral constraints are considered by matching our
expressions to those of one loop $\chi PT$. To account for the medium corrections, the $\rho$ couples to $\pi\pi$ and
$K\bar{K}$ pairs which are properly renormalized in the nuclear medium.}}

\end{abstract}

\begin{center}
\section{Description of the model}
\end{center}
We study the $\rho$ propagation properties by obtaining the $\pi\pi$ and
$K\bar{K}$ scattering amplitudes in the $I=1$ channel. Our states in the
isospin basis are
\ba
\label{isospin}
|\pi\pi>&=&\frac{1}{2}|\pi^+\pi^- - \pi^- \pi^+>\nn \\
|K\overline{K}>&=& \frac{1}{\sqrt{2}} |K^+K^- - K^0\overline{K}^0>.
\ea
Tree level amplitudes
are obtained from the lowest order $\chi PT$ and explicit resonance field
lagrangians of refs. \cite{GyL}, \cite{ecker}. We collect this amplitudes in a
$2\times 2$ $K$ matrix whose elements are

\ba
\label{treeK}
K_{11}(s)&=& \frac{1}{3}\frac{p_{1}^2}{f^2} \left[1+
\frac{2\,G_V^2}{f^2} \frac{s}{M_\rho^2-s}\right]\nn \\
K_{12}(s)&=& \frac{\sqrt{2}}{3}\frac{p_{1} \,p_{2}}{f^2}\left[1+\frac{2\,G_V^2}{f^2}
\frac{s}{M_\rho^2-s}\right] \nn \\
K_{21}(s)&=&K_{12}(s) \nn \\
K_{22}(s)&=& \frac{2}{3}\frac{p_{2}^2}{f^2} \left[1+
\frac{2\,G_V^2}{f^2} \frac{s}{M_\rho^2-s}\right]
\ea
with the labels 1 for $\pi\pi$ and 2 for $K \bar{K}$ states. In equation
(\ref{treeK}) $G_{V}$ is the strength of the pseudoscalar-vector resonance vertex,
$f$ the pion decay constant in the chiral limit, $s$ the squared invariant mass, $M_{\rho}$ the bare mass of the $\rho$ meson and
$p_{i}=\sqrt{s/4-m_{i}^{2}}$.

The final expression of the $T$ matrix is obtained following the N/D method,
which was adapted to the context of chiral theory in ref. \cite{N/D}. It reads
\be
\label{T}
T(s)= \left[I+K(s)\cdot g(s) \right]^{-1}\cdot K(s),
\ee
where $g(s)$ is a diagonal matrix given by the loop of two mesons. In
dimensional regularization it reads

\be
\label{g(s)}
g_i(s)=\frac{1}{16\,\pi^2}\left[-2+d_i+\sigma_{i}(s)\,
\log \frac{\sigma_{i}(s)+1}{\sigma_{i}(s)-1} \right],
\ee
where the subindex $i$ refers to the corresponding two meson state and
$\sigma_{i}(s)=\sqrt{1-4 m_i^2/s}$ with $m_i$ the mass of the particles in the
state $i$. At this stage (vacuum case), the model has proved to be successful
in describing $\pi\pi$ P-wave phase shifts and $\pi$, $K$ electromagnetic
vector form factors \cite{Juan} up to $\sqrt{s}\lesssim
1.2$ GeV. The $d_{i}$ constants contain the information of low energy chiral
constraints. They are obtained by matching the expressions of the form factors
calculated in this approach with those of one-loop $\chi PT$.

In our calculation, in which $g(s)$ is modified in the nuclear medium, we use
cut-off regularization. The $g_{i}(s)$ function with a cut-off in the
three-momentum of the particles in the loop can be found in
Appendix A of ref. \cite{Ramonet}. By comparing the expressions in both
schemes we can get the equivalent $q_i^{max}$ in order to keep the information of
the $d_i$ constants.

\begin{figure}[ht]
\centerline{\includegraphics[width=0.7\textwidth]{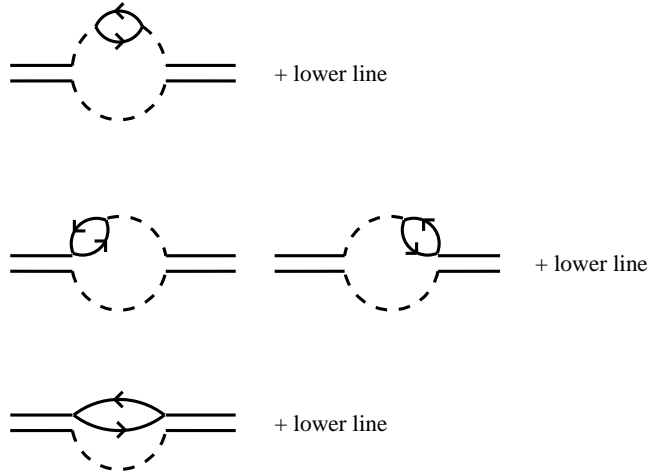}}
\caption{\footnotesize{Medium correction graphs: double solid line represents
the $\rho$
resonance, dashed lines are $\pi$, $K$ mesons in the loop and single solid
lines are reserved for particle-hole
excitations.}}
\end{figure}

Medium corrections are incorporated in the selfenergies of the mesons in the
loops. All the graphs in fig. 1 are considered. In addition to the single $p-h$
bubble we have to account for the medium modifications of the $\rho M M^{'}$
vertex via the $\rho M N$ contact term requested by the gauge invariance of the
theory.

The pion selfenergy is written as usual in terms of the Lindhard
functions. Both $N-h$ and $\Delta -h$ excitations are included. Short range
correlations are also accounted for with the Landau-Migdal parameter $g'$,
set to $0.7$. The final expression is

\be
\label{pionselfenergy}
\Pi_{\pi} (q,\rho)=\vec{q}\,^{2}\, \frac{(\frac{D+F}{2\, f})^{2}\,
U(q,\rho)}{1-(\frac{D+F}{2\, f})^{2}\, g'\, U(q,\rho)}
\ee
where $U=U_N + U_{\Delta}$, the ordinary Lindhard function for $p-h$, $\Delta
-h$ excitations \cite{Chiang}.

The $\bar{K}$ selfenergy has both S-wave and P-wave contributions. The S-wave
piece is obtained from a self-consistent calculation with coupled channels
($\bar{K} N$, $\pi \Sigma$, $\pi \Lambda$, $\eta \Sigma$, $\eta \Lambda$, $K
\Xi$) in which both meson and baryon selfenergies in the medium have been considered.
The P-wave piece includes $\Lambda -h$, $\Sigma -h$ and $\Sigma^{*} (1385)-h$
excitations. The whole $\bar{K}$ selfenergy is borrowed from ref. \cite{Ramos}.

At low energies the $K$ system interacts with nucleons only by S-wave elastic
scattering. We use the expression for the selfenergy from ref. \cite{Ramos2,WKW},

\be
\label{Kselfenergy}
\Pi_{K} (\rho) \simeq 0.13\, m_{K}^{2} \frac{\rho}{\rho_{0}}\, (MeV^{2})
\ee
\begin{figure}[ht]
\centerline{\includegraphics[width=0.7\textwidth]{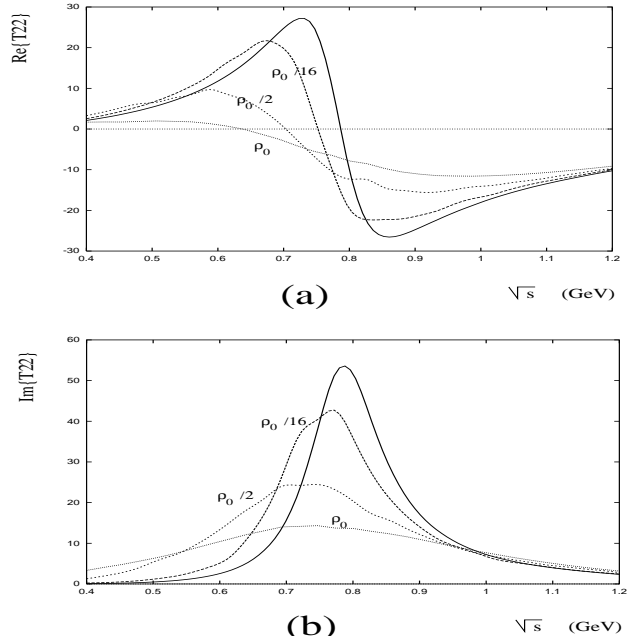}}
\caption{\footnotesize{Real (a) and imaginary (b) parts of the amplitude $T_{\pi
\pi \to \pi \pi}$. The curves are as follows: Solid line, zero density; long dashed line, $\rho=\rho_0/16$;
short dashed line, $\rho=\rho_0/2$; dotted line, normal nuclear density. $\sqrt{s}$ is
the invariant mass of the meson pair.}}
\end{figure}

\section{Results and discussion}

We have plotted in fig. 2 the real and imaginary parts of $T_{22}$ for several
densities. As can be seen, the resonance broadens significantly as density is
increased, its width being around $350$ MeV at $\rho =\rho_{0}$. The zero of
the real part experiences a downward shift which amounts to
$100-150$ MeV at normal density.

Another interesting result comes from the comparison between coupled and decoupled
cases. This is done by setting $K_{12}(s)$ to zero, what automatically makes
the $T$ matrix diagonal. We have found very small differences in $T_{22}$ when
calculating in these two cases. This tells us that in our model the $K \bar{K}$
system has almost no influence on the $\pi \pi \to \pi \pi$ channel even at
normal nuclear density.

\end{document}